\documentclass[preprint,3p]{elsarticle} 
\usepackage{amssymb}
\usepackage{amsfonts}
\usepackage{amsmath}
\usepackage{amsthm}
\usepackage{algorithmic}
\usepackage{algorithm}
\usepackage{booktabs}
\usepackage{url}
\usepackage{multirow}
\usepackage{IEEEtrantools}
\DeclareMathOperator{\cov}{cov}
\usepackage{graphicx,color}
\usepackage[bf,SL,BF]{subfigure}
\usepackage{natbib}\biboptions{authoryear}
\usepackage{color,xcolor}
\usepackage[colorlinks, citecolor=blue, linkcolor=black]{hyperref}
\usepackage{enumerate}
\usepackage[title]{appendix}
\usepackage{lscape}
\usepackage{makecell}
\usepackage{colortbl}
\newcommand{\bS}{\mathbf{S}} \newcommand{\bM}{\mathbf{M}}
 \newcommand{\bX}{\mathbf{X}}
\newcommand{\bR}{\mathbf{R}} 
\newcommand{\bU}{\mathbf{U}}

\newcommand{\bI}{\mathbf{I}}

\newcommand{\cL}{\mathcal{L}}\newcommand{\bLmd}{\boldsymbol{\Lambda}}

\newcommand{\bZ}{\mathbf{Z}}\newcommand{\bSig}{\boldsymbol{\Sigma}}

\newcommand{\bwU}{\widehat{\bU}}
\newcommand{\bwM}{\widehat{\bM}}\newcommand{\bwLmd}{\widehat{\bLmd}}
\newcommand{\bwSig}{\widehat{\bSig}}

\newcommand{\bo}{\mathbf{0}}
\newcommand{\bx}{\mathbf{x}}
 \newcommand{\by}{\mathbf{y}}

\newcommand{\bu}{\mathbf{u}}
\newcommand{\cN}{\mathcal{N}}\newcommand{\cX}{\mathcal{X}} 
\newcommand{\cMN}{\mathcal{MN}} \newcommand{\cM}{\mathcal{M}}\newcommand{\cME}{\mathcal{ME}}
\newcommand{\bw}{\mathbf{w}}

\newcommand{\cW}{\mathcal{W}}

\newcommand\refe[1]{(\ref{#1})}
\newcommand\reff[1]{Figure~\ref{#1}}\newcommand\reft[1]{Table~\ref{#1}}\newcommand\refs[1]{Section~\ref{#1}}

\theoremstyle{plain}  

\def\R{{\mathbb R}}

\newcommand{\tr}{\mbox{$\textrm{\textup{tr}}$}}
\newcommand{\diag}{\mbox{$\textrm{\textup{diag}}$}}

\newcommand{\linspace}{\mbox{$\textrm{\textup{linspace}}$}}

\def\Gam{\mbox{Gam}}
\def\MMD{\mbox{MMD}}

\def \vec{\mbox{vec}}
\def \vec{\mbox{vec}}

\def \exp {\mbox{exp}}

\def\Rmn#1{\expandafter\uppercase\expandafter{\romannumeral #1}}

 \definecolor{purpleliang}{RGB}{148,0,211}

\begin{document}	
		\title{Highly robust factored principal component analysis for matrix-valued outlier accommodation and explainable detection via matrix minimum covariance determinant}
  \begin{frontmatter}
\author[W. Wu,H.Wu]{Wenhui Wu}
\author[W. Wu,C. Shang]{Changchun Shang\corref{cor1}}\ead{scc2017@glut.edu.cn}
\author[J. Zhao]{Jianhua Zhao\corref{cor1}} \ead{jhzhao.ynu@gmail.com}
\author[J. Zhao]{Xuan Ma}
\author[W. Wu,C. Shang]{Yue Wang}

\cortext[cor1]{Corresponding author.}

\address[W. Wu]{School of Mathematics and Statistics, Guilin University of Technology, Guilin, 541006, China}
\address[C. Shang]{Guangxi Colleges and Universities Key Laboratory of Applied Statistics, Guilin University of Technology, Guilin, 541006, China}
\address[J. Zhao]{School of Statistics and Mathematics, Yunnan University of Finance and Economics, Kunming, 650221, China}
\address[H.Wu]{School of Mathematics, College of Science, Chang'an University, Xi'an, 710064, China}

\begin{abstract}
		Principal component analysis (PCA) is a classical and widely used method for dimensionality reduction, with applications in data compression, computer vision, pattern recognition, and signal processing. However, PCA is designed for vector-valued data and encounters two major challenges when applied to matrix-valued data with heavy-tailed distributions or outliers: (1) vectorization disrupts the intrinsic matrix structure, leading to information loss and the curse of dimensionality, and (2) PCA is highly sensitive to outliers. Factored PCA (FPCA) addresses the first issue through probabilistic modeling, using a matrix normal distribution that explicitly represents row and column covariances via a separable covariance structure, thereby preserving the two-way dependency and matrix form of the data. Building on FPCA, we propose highly robust FPCA (HRFPCA), a robust extension that replaces maximum likelihood estimators with the matrix minimum covariance determinant (MMCD) estimators. This modification enables HRFPCA to retain FPCA’s ability to model matrix-valued data while achieving a breakdown point close to 50\%, substantially improving resistance to outliers. Furthermore, HRFPCA produces the score–orthogonal distance analysis (SODA) plot, which effectively visualizes and classifies matrix-valued outliers. Extensive simulations and real-data analyses demonstrate that HRFPCA consistently outperforms competing methods in robustness and outlier detection, underscoring its effectiveness and broad applicability.
\end{abstract}
	
	
	\end{frontmatter}
	\section{Introduction}\label{sec:intr} 
    Principal component analysis (PCA) \citep{Jolliffe2002} is a classical and widely adopted method for dimensionality reduction that extracts the principal directions of variation by performing eigenvalue decomposition of the sample covariance matrix, thereby maximizing information retention and supporting diverse applications in data compression, computer vision, pattern recognition, and signal processing. Despite its popularity, PCA exhibits two critical limitations in modern applications: (1) it is inherently designed for vector-valued data, so vectorizing matrix-valued observations can disrupt intrinsic row–column structures and exacerbate the curse of dimensionality; and (2) it is highly sensitive to outliers, as the sample covariance matrix is easily distorted by heavy-tailed or contaminated data, producing principal directions that may deviate from the underlying variation of typical observations.

To address the first limitation, numerous dimensionality reduction methods have been developed to operate directly on matrix-valued data, preserving its intrinsic two-dimensional structure. For instance, two-dimensional PCA \citep{zhangdq-2dpca} and generalized low-rank approximation of matrices \citep{jpye-glram} exploit the row–column structure of matrices, mitigating the curse of dimensionality and substantially reducing computational complexity. In contrast to these non-probabilistic approaches, factored principal component analysis (FPCA) \citep{sbpca-dryden} employs a probabilistic modeling framework based on the matrix normal distribution, explicitly representing row and column covariances via a separable structure to preserve the matrix form. FPCA uses a `\textit{flip-flop}' algorithm to obtain maximum likelihood (ML) estimates and has demonstrated superior performance on various face recognition datasets, inspiring the development of subsequent statistical methods and applications \citep{John2012-Evaluating, zhao2012-slda, zhao2012-bppca, Paromita2020, Thompson2020, Bhargab2022, zhao2023-rfpca}. Nevertheless, FPCA remains sensitive to heavy-tailed distributions or contaminated data, limiting its applicability in the presence of atypical or outlying observations.

To address the second limitation, several robust PCA variants have been proposed to derive principal components resilient to outliers, often by replacing the classical covariances with robust estimates. For example, \cite{Maronna1976-robust} introduced affine-equivariant M-estimators of multivariate scatter. However, \cite{Dumbgen-bkd} showed that the M-functional of the multivariate $t$ distribution—equivalent to the ML estimator for $\nu>1$—has a theoretical breakdown point of $1/(p+\nu) < 1/p$, where $p$ is the data dimension and $\nu$ the degrees of freedom. This implies that such methods tolerate only a very small fraction of outliers; e.g., for $p=1000$, they can handle less than 0.1\% contamination, severely limiting practical applicability. A far more robust alternative is the Minimum Covariance Determinant (MCD) estimator \citep{Rousseeuw1984-mcd}. When the subset size is set to $h = \lfloor (n + p + 1)/2 \rfloor$, the MCD achieves a maximum breakdown point of $(n-h+1)/n$, approaching 50\% as $n$ increases. While robust PCA methods based on the MCD \citep{croux2000-principal, Hubert2005-robpca} inherit this strong robustness, they are inherently restricted to vector-valued data.

To jointly address both limitations, \cite{Thompson2020} and \cite{zhao2023-rfpca} respectively proposed TPCA and RFPCA as natural extensions of FPCA, each grounded in a distinct formulation of the matrix $t$-distributions, as presented in equations \refe{eqn:T.pdf} and \refe{eqn:mvt.pdf}. Although their theoretical breakdown points have not been formally established, empirical evidence suggests that they exceed the theoretical breakdown point $1/(p+\nu)$ of the multivariate $t$-distribution, yet remain well below the maximum breakdown point achievable by MCD. More recently, \cite{Mayrhofer2025-mmcd} extended the MCD framework to matrix-valued data, introducing the matrix minimum covariance determinant (MMCD) estimator with a maximum breakdown point of $\lfloor (n-d)/2 \rfloor / n$ for $h = \lfloor (n+d+2)/2 \rfloor$, where $d = \lfloor d_r/d_c + d_c/d_r \rfloor$. As $n \to \infty$, the breakdown point approaches 50\%, reflecting high robustness. Motivated by the strengths of FPCA in modeling matrix-valued data and the superior robustness of MMCD, we develop a novel, highly robust PCA framework, with the main contributions summarized as follows:
\begin{enumerate}[(i)]
	\item We propose HRFPCA, a highly robust extension of FPCA incorporating the MMCD estimators. HRFPCA combines FPCA’s capability for modeling matrix-valued data with the high robustness of MMCD, overcoming the limitations of vectorization and achieving a theoretical breakdown point approaching 50\%, substantially exceeding that of existing methods such as RFPCA and TPCA.
	\item We introduce a score–orthogonal distance analysis (SODA) plot tailored for matrix-valued data, enabling effective visualization and classification of different types of outliers.
\end{enumerate}

The remainder of this paper is organized as follows. In \refs{sec:reworks}, we review related methods. In \refs{sec:hrfpca}, we present HRFPCA and the SODA plot. \refs{sec:syn.expr} and \refs{sec:real.expr} report extensive experiments on synthetic and real-world data, respectively, comparing HRFPCA with existing methods. Finally, in \refs{sec:discussion}, we conclude and discuss directions for future research

\section{Related works}\label{sec:reworks}
This section provides a brief review of several methods relevant to this work, including FPCA, TPCA, RFPCA, and MMCD.
\subsection{FPCA}\label{sec:fpca}
Assume a random matrix $\bX \in \R^{d_c \times d_r}$ follows a matrix normal distribution, with probability density function
\begin{equation}\label{eqn:mvn}
	p(\bX) = (2\pi)^{-\frac{d_c d_r}{2}}|\bSig_c|^{-\frac{d_r}{2}}|\bSig_r|^{-\frac{d_c}{2}}\exp\left\{-\frac{1}{2}\MMD^2(\bX)\right\}
\end{equation}
where $\bM$ denotes the mean matrix, $\bSig_c$ and $\bSig_r$ represent the column and row covariance matrices, respectively, and $\MMD^2(\bX)$ is the squared matrix Mahalanobis distance:
\begin{equation}\label{eqn:mat.mahlanobis}
	\MMD^2(\bX) = \tr\left\{\bSig_c^{-1}(\bX-\bM)\bSig_r^{-1}(\bX-\bM)^T\right\}.
\end{equation}
For convenience, the distribution in \refe{eqn:mvn} is abbreviated as $\bX \sim \cMN_{d_c, d_r}(\bM, \bSig_c, \bSig_r)$.

Consider an orthogonal transformation $\by=\bU'\bx$, where $\bU \in \R^{d \times q}$. The covariance matrix of $\by$ is then $\bU'\bSig\bU$, and the goal of PCA is to identify $\bU$ that maximizes the variance in the projected space:
\begin{equation}\label{eqn:pca}
	\mathop{\hbox{max}}_\bU\hbox{tr}\left\{\bU^T\bSig\bU\right\},\quad s.t. \quad \bU^T\bU=\bI_q.
\end{equation}

Under the matrix normal assumption, the vectorized form satisfies $\bx = \vec(\bX) \sim \cN_{d_cd_r}(\vec(\bM), \bSig_r \otimes \bSig_c)$, where $\otimes$ denotes the Kronecker product. Consequently, the covariance matrix factorizes as $\bSig = \bSig_r \otimes \bSig_c$, and the projection matrix can be expressed as $\bU_r \otimes \bU_c$. Substituting this factorization into \eqref{eqn:pca} decomposes the optimization into two subproblems, each resembling standard PCA:
\begin{equation}\label{eqn:fpca}
	\mathop{\hbox{max}}_{\bU_c}\hbox{tr}\left\{\bU^T_c\bSig_c\bU_c\right\},\,\,s.t.\ \, \bU^T_c\bU_c=\bI_{q_c},\quad \mbox{and}\quad \mathop{\hbox{max}}_{\bU_r}\hbox{tr}\left\{\bU^T_r\bSig_r\bU_r\right\},\,\,s.t.\ \, \bU^T_r\bU_r=\bI_{q_r}.
\end{equation}
This factorization motivates the term factored principal component analysis (FPCA). The matrices $\bU_c$ or $\bU_r$ are determined up to orthogonal rotations of sizes $q_c \times q_c$ and $q_r \times q_r$, respectively. Let $\{(\bu_{ci},\lambda_{ci})\}_{i=1}^{q_c}$ and $\{(\bu_{ri},\lambda_{ri})\}_{i=1}^{q_r}$ denote the descending eigen-pairs of $\bSig_c$ and $\bSig_r$, and let $\bLmd_c=\diag\{\lambda_{c1},\lambda_{c2},\dots,\lambda_{cq_c}\}$ and $\bLmd_r=\diag\{\lambda_{r1},\lambda_{r2},\dots,\lambda_{rq_r}\}$ be the corresponding diagonal matrices of eigenvalues. Then, the principal components (PCs) are given by
\begin{equation}\label{eqn:fpca.U}
	\bU_c=[\bu_{c1},\dots,\bu_{cq_c}],\,\,\bU_r=[\bu_{r1},\dots,\bu_{rq_r}],\, \hbox{top}\ q_c,q_r\ \hbox{eigenvectors of}\ \bSig_c\ \hbox{and}\ \bSig_r.
\end{equation}

Given $n$ independent and identically distributed (i.i.d.) samples $\{\bX\}_{i=1}^{n}$, the ML estimates are obtained by maximizing the log-likelihood function and alternately updating \refe{eqn:Sigc.est} - \refe{eqn:Sigr.est} using the `\textit{flip-flop}' algorithm \citep{Dutilleul1999},
\begin{IEEEeqnarray}{rCl}
	\bwM &=& \frac{1}{n} \sum_{i=1}^{n}\bX_i, \label{eqn:M.est}\\
	\bwSig_c &=& \frac{1}{nd_r}\sum_{i=1}^{n}(\bX_i - \bwM)\bwSig_r^{-1}(\bX_i - \bwM)^T, \label{eqn:Sigc.est}\\
	\bwSig_r &=& \frac{1}{nd_c}\sum_{i=1}^{n}(\bX_i - \bwM)^T\bwSig_c^{-1}(\bX_i - \bwM),  \label{eqn:Sigr.est}
\end{IEEEeqnarray}

Given the number of PCs $(q_c, q_r)$, the corresponding factorized PCs, $\bU_c$ and $\bU_r$, can be computed directly from \refe{eqn:fpca.U}.

\subsection{TPCA}\label{sec:TPCA}
Let $\bS\sim\cW_{d_c}(\bSig_c^{-1},\nu+c-1)$ be a latent random matrix, and assume that conditional on $\bS$, $\bX|\bS \sim \cMN_{d_c,d_r}(\bM, \bS^{-1}, \bSig_r)$. The marginal probability density function of $\bX$ is then obtained by integrating out $\bS$:
\begin{IEEEeqnarray}{rCl}\label{eqn:T.pdf}
	p\left(\bX\right)&=& \frac{\left|\bSig_r\right|^{-\frac{d_c}{2}}\left|\bSig_c\right|^{-\frac{d_r}{2}} \Gamma_{c}\left(\frac{\nu+d_c+d_r-1}{2}\right)}{(\pi)^{\frac{d_c d_r}{2}} \Gamma_{c}\left(\frac{\nu+d_c-1}{2}\right)} \times\left|\bI_{c}+\bSig_c^{-1}(\bX-\bM) \bSig_r^{-1}(\bX-\bM)^T\right|^{\frac{\nu+d_c+d_r-1}{2}}, 
\end{IEEEeqnarray}
where $\bM \in \R^{d_c \times d_r}$ is the location matrix, $\nu$ is the degrees of freedom, and $\bSig_c$ and $\bSig_r$ are positive definite matrices. Following \citet{zhao2023-rfpca}, this distribution is referred to as the matrix $T$-distribution, denoted $\bX \sim \cM xT_{d_c, d_r}(\bM, \bSig_c, \bSig_r, \nu)$. It can be shown that the covariance of $\bX$ \citep{gupta-mvn} is
\begin{equation}\label{eqn:T.covx}
	\cov{(\vec(\bX))}=\frac1{\nu-2}\bSig_r\otimes\bSig_c, \quad n > 2.
\end{equation}

Analogous to FPCA, one can ignore the constant factor $1 / (\nu -2 )$ in \refe{eqn:T.covx} and substitute the separable covariance structure $\bSig_r \otimes \bSig_c$ into the PCA optimization problem \eqref{eqn:pca}. This leads to the same two subproblems in \eqref{eqn:fpca}, whose solutions $\bU_c$ and $\bU_r$ are obtained as in \eqref{eqn:fpca.U}. This approach is referred to as TPCA. The matrices $\bSig_c$ and $\bSig_r$ can be estimated via ML method \citep{Thompson2020}.

\subsection{RFPCA}\label{sec:rfpca}
Let $\tau \sim \Gam(\nu/2,\nu/2)$ be a latent variable, and assume that conditional on $\tau$, the observation matrix $\bX|\tau \sim \cMN_{d_c,d_r}(\bM,\bSig_c/\tau,\bSig_r)$. The marginal probability density function of $\bX$ is then given by
\begin{equation}\label{eqn:mvt.pdf}
	p(\bX)=\frac{|\bSig_c|^{-\frac{d_r}{2}}|\bSig_r|^{-\frac{d_c}{2}} \Gamma(\frac{\nu+d_cd_r}{2})}{(\pi \nu)^{\frac{d_cd_r}{2}}\Gamma(\frac{\nu}{2})}\left(1+\frac{1}{\nu}\MMD^2(\bX)\right)^{-\frac{\nu+d_cd_r}{2}},
\end{equation}
where $\bM$, $\bSig_c$ and $\bSig_r$ are as defined in \refe{eqn:mvn}, and $\nu$ denotes the degrees of freedom parameter. Following  \citet{zhao2023-rfpca}, this distribution is referred to as the matrix $t$-distribution, abbreviated as $\bX \sim \cM t_{d_c, d_r}(\bM, \bSig_c, \bSig_r, \nu)$. Its vectorized form satisfies a multivariate $t$-distribution \citep{Liu1995}, i.e., $\vec(\bX)\sim t_{d_cd_r}(\vec(\bM), \bSig_r\otimes\bSig_c,\nu)$, and its covariance is
\begin{equation}
	\cov{(\vec(\bX))}=\frac{\nu}{\nu-2}\bSig_r\otimes\bSig_c.\label{eqn:mvt.covx}
\end{equation} 

Under the matrix $t$-distribution, analogous to FPCA, one can ignore the constant factor $\nu/(\nu-2)$ in \refe{eqn:mvt.covx} and substitute the separable covariance $\bSig_r \otimes \bSig_c$ into the PCA optimization problem \refe{eqn:pca}. This yields the same two sub-optimization problems in \refe{eqn:fpca}, with solutions $\bU_c$ and $\bU_r$ as in \refe{eqn:fpca.U}. This approach is referred to as RFPCA, where $\bSig_c$ and $\bSig_r$ in \refe{eqn:mvt.pdf} can be estimated via ML method using an EM-type algorithm \citep{zhao2023-rfpca}.

FPCA assumes that the data follow a matrix normal distribution, which can be sensitive to heavy tails or outliers. In contrast, TPCA and RFPCA are based on two types of matrix $t$-distributions (\refe{eqn:T.pdf} and \refe{eqn:mvt.pdf}), each incorporating a degrees-of-freedom parameter $\nu$ that allows the model to accommodate heavy-tailed or outlier-prone data. Therefore, TPCA and RFPCA are more robust than FPCA. However, the theoretical breakdown points of these two methods have not yet been established; \cite{zhao2023-rfpca} only provides an empirical assessment of their robustness, which exceeds the theoretical breakdown point derived from the multivariate $t$-distribution. Nevertheless, compared to the high robustness of MCD (maximum theoretical breakdown point close to 50\%), a gap remains.

\subsection{MMCD}\label{sec:mmcd}
In robust statistics, the MCD estimator is one of the most widely used robust estimators due to its high breakdown point. \cite{Raymaekers2024-cellwise} demonstrated that the MCD estimator can be reformulated within a likelihood framework. The objective of the MCD estimator is to identify a subset of $h$ observations from $n$ samples ($n/2 \leq h \leq n$) such that the determinant of the corresponding sample covariance matrix is minimized. This is equivalent to selecting the $h$-subset that maximizes the multivariate normal log-likelihood function. However, MCD is designed for vector-valued data, and its direct application to matrix-valued data suffers from the curse of dimensionality.

To address this limitation, \cite{Mayrhofer2025-mmcd} extended the idea of multivariate MCD to the family of matrix elliptical distributions \citep{gupta2012-Elliptically}. Suppose the probability density function of a random matrix $\bX$ can be written as
\begin{eqnarray}
	\nonumber
	p(\bX) &=& \left|\bSig_r\right|^{-\frac{d_c}{2}}\left|\bSig_c\right|^{-\frac{d_r}{2}}g\left(\tr\left\{\bSig_c^{-1}(\bX-\bM)\bSig_r^{-1}(\bX-\bM)^T\right\}\right) \\
	&=& \left|\bSig_r\right|^{-\frac{d_c}{2}}\left|\bSig_c\right|^{-\frac{d_r}{2}}g\left(\MMD^2(\bX)\right),
\end{eqnarray} 
then $\bX$ follows a matrix elliptical distribution, denoted as $\cME(\bM, \bSig_c, \bSig_r, g)$, where $g: [0,\infty) \to \R$ is the density generator function. It is evident that, for a suitable choice of $g$, the three distributions \refe{eqn:mvn}, \refe{eqn:T.pdf}, and \refe{eqn:mvt.pdf} assumed by FPCA, TPCA, and RFPCA, respectively, are all special cases of the matrix elliptical family.

Consider $n$ i.i.d. samples $\cX = \{\bX\}_{i=1}^{n}$ drawn from a matrix normal distribution $\cMN_{d_c,d_r}(\bM, \bSig_c, \bSig_r)$. Introducing a weight vector $\bw = (w_1, w_2, \cdots, w_n) \in \R^n$, The weighted log-likelihood function is given by
\begin{IEEEeqnarray}{rCl}\label{eqn:weight.logL}
	\cL(\bw,\bM,\bSig_c, \bSig_r \mid \cX) = -\frac{1}{2}\sum_{i=1}^{n}w_i\left(d_r\ln|\bSig_c|+d_c\ln|\bSig_r|+\MMD^2(\bX)+d_cd_r\ln 2\pi\right),
\end{IEEEeqnarray}
When $w_i = 1$, for all $i = 1, \cdots, n$, equation \refe{eqn:weight.logL} reduces to the standard log-likelihood function of the matrix normal distribution, whose maximization yields the classical ML estimates in equations \refe{eqn:M.est}-\refe{eqn:Sigr.est}. By introducing binary weights $w_i \in \{0,1\}$ with the constraint $\sum_{i=1}^{n}w_i = h$, one effectively trims $n-h$ observations. Since the contribution from outliers should be excluded, the goal is to identify a subset $H \in \{1,2,\cdots,n\}$ of size $|H| = h$ such that $w_i = 1$ if $i \in H$ and $w_i = 0$ otherwise. The corresponding constrained optimization problem for the weighted ML estimates is 
\begin{IEEEeqnarray}{rCl}\label{eqn:max.w.logL}
	\nonumber
	& \mathop{\hbox{max}}_{\bw,\bM,\bSig_c,\bSig_r} \cL(\bw,\bM,\bSig_c, \bSig_r \mid \cX) \\
	& \text{s.t.} \quad w_i \in \{0,1\} \ \text{for all}\ i =1,\cdots, n \quad \text{and} \quad \sum_{i=1}^{n} w_i = h.
\end{IEEEeqnarray}
When $n/2 < h < n$, $h \geq \lfloor d_c/d_r + d_r/d_c \rfloor + 2$, maximizing \refe{eqn:max.w.logL} over all $h$-subsets $H$ is equivalent to minimizing
\begin{IEEEeqnarray}{rCl}\label{eqn:min.w.logL}
	\ln|\bSig_r^H \otimes \bSig_c^H| = d_c\ln|\bSig_r^H| + d_r \ln|\bSig_c^H|,
\end{IEEEeqnarray}
yielding the ML estimates based on the subset $H$:
\begin{IEEEeqnarray}{rCl}
	\bwM^H &=& \frac{1}{h} \sum_{i \in H} \bX_i, \label{eqn:M.est.h}\\
	\bwSig_c^H &=& \frac{1}{hd_r}\sum_{i \in H}\left(\bX_i - \bwM^H\right)\bwSig_r^{-1,H}\left(\bX_i - \bwM^H\right)^T, \label{eqn:Sigc.est.h}\\
	\bwSig_r^H &=& \frac{1}{hd_c}\sum_{i \in H}\left(\bX_i - \bwM^H\right)^T\bwSig_c^{-1,H}\left(\bX_i - \bwM^H\right),  \label{eqn:Sigr.est.h}
\end{IEEEeqnarray}
where $\bwSig_c^{H,-1}$ and $\bwSig_r^{H,-1}$ denote the inverses of $\bwSig_c^H$ and $\bwSig_r^H$, respectively. Consequently, given $n$ i.i.d. samples from a continuous matrix-variate distribution, \cite{Mayrhofer2025-mmcd} defined the estimators that minimizes \refe{eqn:min.w.logL} as the raw MMCD estimators. In the special cases where $d_c = 1$ or (and) $d_r = 1$, the optimization problem \refe{eqn:min.w.logL} reduces to the classical MCD optimization, yielding the multivariate (or univariate) MCD estimators, respectively.

Based on the properties of matrix elliptical distributions, \cite{Mayrhofer2025-mmcd} established that the MMCD estimators are equivariant under matrix affine transformations. When $h = \lfloor (n + d + 2) \rfloor/2$, its maximum breakdown point is given by $\lfloor (n - d + 1)/2 \rfloor / n$, where $d = \lfloor d_c/d_r + d_r/d_c \rfloor$. Both theoretical analysis and empirical studies demonstrate that the MMCD estimators achieve a higher breakdown point for matrix data than the MCD estimators applied to vectorized data. Furthermore, a computationally efficient algorithm, analogous to the Fast-MCD algorithm \citep{Rousseeuw1999-FastMCD}, was developed to compute the MMCD estimators in practice.

\subsubsection{Explainable outlier detection}\label{sec:out.dect}
Assume that the dataset $\{\bX\}_{i=1}^{n}$ comes from a matrix normal distribution. The objective of outlier detection is to identify observations far from the data center.  In robust statistics, the squared Mahalanobis distance is widely employed for this purpose. Specifically, an observation $\bX_i$ is considered as an outlier if
\begin{equation}\label{eqn:out.dect}
	\MMD^2\left(\bX_i; \bwM, \bwSig_c, \bwSig_r\right) > \chi_{d_cd_r; 0.975}^2,\quad i = 1, \cdots, n,
\end{equation}
where $\bwM, \bwSig_c, \bwSig_r$ represent robust estimates \citep{Mayrhofer2025-mmcd}. 

RFPCA also provides an alternative approach to outlier detection by utilizing the posterior expectation of the latent variable $\tau$ \citep{zhao2023-rfpca}. However, this method is inherently model-specific and may fail when the data do not satisfy the assumptions underlying RFPCA. In contrast, the squared Mahalanobis distance in \refe{eqn:out.dect} offers a more general diagnostic tool. It is model-agnostic and requires only the substitution of robust estimates of $\bwM$, $\bwSig_c$ and $\bwSig_r$, such as the ML estimates from TPCA or RFPCA or the MMCD estimates defined in equations \refe{eqn:M.est.h} - \refe{eqn:Sigr.est.h}. 

Although  the squared Mahalanobis distance in \refe{eqn:out.dect} is very useful for identifying outliers, it does not provide insight into the underlying causes of these anomalies. To address this limitation and enable a more detailed interpretation of the outlier, \cite{Mayrhofer2025-mmcd} extended the concept of multivariate Shapley values \citep{Mayrhofer2023-multivariate} to the setting of matrix-valued data. This extension allows for the decomposition of the outlyingness into cellwise contributions, thereby offering a clearer and more granular understanding of the anomaly. The contribution of each element of matrix $\bX$ is computed as
\begin{equation}\label{eqn:cellwise}
	\Phi(\bX) = (\bX - \bM)\circ\bSig_{c}^{-1}(\bX - \bM)\bSig_{r}^{-1},
\end{equation}
where $\circ$ denotes element-wise multiplication. For details on computing row-wise and column-wise Shapley values, see \cite{Mayrhofer2025-mmcd}.

\section{Methodology}\label{sec:hrfpca}
\subsection{The proposed highly robust factored PCA (HRFPCA)}\label{sec:desc}
As mentioned in \refs{sec:mmcd}, when the observations follow a matrix normal distribution, the MMCD estimators are obtained by maximizing the weighted log-likelihood \refe{eqn:max.w.logL} over all subsets $H \in \{1,2,\cdots,n\}$ of size $|H| = h$. This implies that, within the clean $h$-subset, the ML estimates derived from the log-likelihood function in \refe{eqn:mvn} coincide with the MMCD estimates. Consequently, substituting the MMCD estimates $\bwSig_c^H$ and $\bwSig_r^H$ into the optimization problem \refe{eqn:fpca} yields a version of FPCA with a high breakdown point, which we refer to as HRFPCA. Specifically,
\begin{equation}\label{eqn:hrfpca}
	\bwSig^H = \bwSig_r^H \otimes \bwSig_c^H = \left(\bwU_r^H\bwLmd_r^H\bwU_r^{H,T}\right)\otimes\left(\bwU_c^H\bwLmd_c^H\bwU_c^{H,T}\right),
\end{equation}
where $\bwU_c^H \in \R^{d_c \times q_c}$, $\bwU_r^H \in \R^{d_r \times q_r}$, and $\bwLmd_c^H$ and $\bwLmd_r^H$ are defined as in \refs{sec:fpca}, yielding $\bwSig_c^H$ and $\bwSig_r^H$. Based on the projection matrices $\bwU_c^H$ and $\bwU_r^H$, the robust principal component score matrix $\bZ_i^H$ is computed as:
\begin{equation}\label{eqn:hrfpca.Z}
	\bZ_i^H = \bwU_c^{H,T}\left(\bX_i - \bwM^H\right)\bwU_r^H,
\end{equation}
where, given the PCs number $q_c$ and $q_r$, it follows that $\bZ_i^H \in \R^{q_c \times q_r}$.

HRFPCA can thus be viewed as a two-stage procedure: in the first stage, a fast iterative reweighted MMCD algorithm is applied to obtain robust estimates of parameters $\bM$, $\bSig_c$ and $\bSig_r$; in the second stage, FPCA is performed using these robust covariance estimates. The full procedure is summarized in Algorithm \ref{alg:hrfpca}.
\begin{algorithm}[htb]
	\caption{HRFPCA procedure.}
	\label{alg:hrfpca}
	\begin{algorithmic}[1]
		\REQUIRE Data $\cX = (\bX_1,\cdots, \bX_n)$, the number of PCs $(q_c, q_r)$.
		\STATE \emph{Stage 1:} Obtaining the clean subset $\cX^H$ and the corresponding MMCD estimators $\bwM^H$, $\bwSig_c^H$, and $\bwSig_r^H$ are achieved via the fast reweighted MMCD procedure, i.e., Algorithm 2 in the supplementary material of \cite{Mayrhofer2025-mmcd}.
		\STATE \emph{Stage 2:} Calculating the robust column PCs $\bwU_c^H \in \R^{d_c \times q_c}$ and the row PCs $\bwU_r^H \in \R^{d_r \times q_r}$ via \refe{eqn:hrfpca}, respectively, and calculating the matrix-valued scores $\bZ_i^H$'s via \refe{eqn:hrfpca.Z}, $i = 1,\cdots, n$.
		\ENSURE $\cX^H$, $\bwM^H$, $\bwSig_c^H$, $\bwSig_r^H$, $\bwU_c^H$, $\bwU_r^H$, and $\{\bZ_i^H\}_{i=1}^n$.
	\end{algorithmic}
\end{algorithm}

\subsection{The score–orthogonal distance analysis (SODA) plot}\label{sec:diag.plot}
As emphasized by \cite{Hubert2005-robpca}, robust PCA serves two primary objectives: (1) to identify the linear combinations of the original variables that capture most of the data variability, even in the presence of outliers; and (2) to detect outliers and determine their type. The first objective has been accomplished in \refs{sec:desc}; this section focuses on the second. Following the framework in \cite{Hubert2005-robpca, zhao2023-rfpca}, observations can be categorized into four types: regular observations, good leverage points (outlying \textit{only} in the PC subspace), orthogonal outliers (outlying \textit{only} in the OC subspace), and bad leverage points (outlying in \textit{both} the PC and OC subspaces). Note that both OC and PC+OC outliers are classified simply as OC outliers in \cite{she2016robust}. However, experimental results in \refs{sec:outflag.syn.expr} demonstrate that although both OC and PC+OC outliers correspond to bad leverage points, a high proportion of PC outliers can also substantially distort the subspace estimation. 

To more accurately distinguish regular observations from the other three types of outliers in matrix-valued data, we construct an outlier diagnostic plot adapted for matrix-valued observations, following \cite{Hubert2005-robpca}. On the horizontal axis, we plot the robust score distance ($SD_i$) for each matrix observation, 
\begin{equation}\label{eqn:SD}
	SD_i = \sqrt{\sum_{k=1}^{q_c}\sum_{j=1}^{q_r}\frac{z_{kj}^2}{\lambda_{ck}\lambda_{rj}}},
\end{equation}
where $z_{kj}$, computed from \refe{eqn:hrfpca.Z}, denotes the element in the $k$-th row and $j$-th column of the matrix $\bZ_i$. On the vertical axis, we plot the orthogonal distance of each matrix observation to the PC subspace, defined as
\begin{equation}
	OD_i = \tr\left\{\bR_i^T\bR_i\right\},   \label{eqn:OD} 
\end{equation}
where $\bR_i = \bX_i -\bwM^H - \bwU_c^H\bZ_i^H\bwU_r^{H,T}$. Since both axes represent distances, we refer to this plot as the score–orthogonal distance analysis (SODA) plot.

The SODA plot offers an intuitive diagnostic: observations that deviate substantially from the subspace `\textit{bubble up}', analogous to soda bubbles rising to the surface, thereby allowing outliers to be visually distinguished from regular observations. Two threshold lines are drawn to identify different types of outliers. The horizontal threshold is based on the robust score distance, which, under a normal distribution, follows a squared Mahalanobis distance asymptotically distributed as $\chi_{q_cq_r}^2$. Therefore, the horizontal threshold is set to $\sqrt{\chi_{q_cq_r; 0.975}^2}$. Determining the vertical threshold is more challenging, as the distribution of the orthogonal distances is unknown. A scaled chi-squared distribution $c_1\chi_{c_2}^2$ provides a reasonable approximation of the unknown distribution of the squared orthogonal distances, with parameters $c_1$ and $c_2$ estimated via the method of moments. Following \cite{Hubert2005-robpca}, we apply the Wilson-Hilferty approximation, implying that $OD_i^{2/3}$ approximately follows a normal distribution with mean $\mu = (c_1c_2)^{1/3}(1-2/(9c_2))$ and variance $\sigma^2 = 2c_1^{2/3}/(9c_2^{1/3})$. The mean $\hat{\mu}$ and variance $\hat{\sigma}^2$ are estimated using the univariate MCD method. The vertical threshold is then defined as $(\hat{\mu} + \hat{\sigma} z_{0.975})^{3/2}$, where $z_{0.975} = \Phi^{-1}(0.975)$ corresponds to the $97.5\%$ quantile of the standard normal distribution.

By combining both axes, the SODA plot enables a clear categorization of outlier types in matrix-valued data: observations exceeding only the horizontal threshold are typically good leverage points, those exceeding only the vertical threshold are orthogonal outliers, and observations exceeding both thresholds correspond to bad leverage points. In this way, the `soda bubbles' analogy helps visualize how outliers rise above the cloud of regular observations, making them easy to detect and interpret.

\section{ Simulations}\label{sec:syn.expr}
This section investigates the performance of the proposed HRFPCA method and the SODA plot using synthetic data. For clarity of exposition, define $\bu_1 = [1/\sqrt{2}, -1/\sqrt{2}, 0, \dots, 0]'$, $\bu_2 = [0, 0, 1/\sqrt{2}, -1/\sqrt{2}, 0, \dots, 0]'$, and $\bu_3 = [0, 0, 0, 0, 1/\sqrt{2}, -1/\sqrt{2}, 0, \dots, 0]'$, and generate the following two datasets.
\begin{enumerate}[(i)]
	\item \emph{Data1}: \emph{Data1} is generated from a matrix normal distribution $\cMN_{4,10}(\bM, \bSig_c, \bSig_r) $, with sample size $n = 1000$, $\bM=\bo$. The eigenvalues of $\bSig_{c}$ and $\bSig_{r}$ are $(5,\linspace(0.8,0.5,3))$ and $(4,3,2,\linspace(0.5,0.3,7))$, respectively. The PCs of $\bSig_{c}$ and $\bSig_{r}$ are $\bU_c = \bu_1$ and $\bU_r = (\bu_1, \bu_2, \bu_3)$, respectively. Since the first eigenvalue of $\bSig_{c}$ and the first three eigenvalues of $\bSig_{r}$ dominate, the numbers of PCs $q_c$ and $q_r$ are 1 and 3, respectively.
	\item \emph{Data2}: \emph{Data2} is generated from a matrix normal distribution $\cMN_{10,8}(\bM, \bSig_c, \bSig_r)$ with sample size $n=200$ and $\bM=\mathbf{5}$. The eigenvalues of $\bSig_c$ and $\bSig_r$ are $(5,4,3,\linspace(0.8,0.5,7))$ and $(4,3,\linspace(0.5,0.3,6))$, respectively. The PCs of $\bSig_{c}$ and $\bSig_{r}$ are $\bU_c = (\bu_1, \bu_2, \bu_3)$ and $\bU_r = (\bu_1, \bu_2)$, respectively. Accordingly, the numbers of PCs $q_c$ and $q_r$ are 3 and 2, respectively.
\end{enumerate}

 For HRFPCA, the size of the `clean' subset is $h = \lfloor(n + d + 2)/2 \rfloor$, where $d = \lfloor d_c/d_r + d_r/d_c \rfloor$. For the iterative algorithms for paramater estimation of FPCA, TPCA, and RFPCA, random initialization is employed, and the iteration is terminated when the relative change in the objective function, ($ |1 - \cL^{(t)} / \cL^{(t+1)} | $), falls below a threshold ($tol = 10^{-8}$) or when the number of iterations exceeds $t_{max} = 1000$. Unless otherwise specified, the settings described above are used throughout the subsequent experiments.

\subsection{Robustness of HRFPCA}\label{sec:hrfpca.syn.expr}
This experiment evaluates the robustness of HRFPCA by comparing its performance with three related matrix-type PCA methods -- FPCA \citep{sbpca-dryden}, TPCA \citep{Thompson2020}, and RFPCA \citep{zhao2023-rfpca} -- using synthetic data contaminated with outliers. Specifically, in \emph{Data1}, $np$ normal observations are replaced with outliers, and the analysis is conducted on the resulting contaminated dataset. Five contamination proportions are considered: $p = 0.1, 0.3, 0.4, 0.49,$ and $0.5$. Following the framework in \cite{zhao2023-rfpca}, we examine four distinct scenarios for generating three types of outliers, yielding a total of 12 outlying cases. The three types of outliers, defined in \refs{sec:diag.plot}, include PC outliers, OC outliers, and PC+OC outliers. These outliers are generated under the following four scenarios: Sit-\Rmn1: $\bU(-100, 100)$, Sit-\Rmn2: $\bU(-10000, 10000)$, Sit-\Rmn3: $\bU(100, 110)$, and Sit-\Rmn4: $\bU(10000, 11000)$, where $\bU(a,b)$ denotes the uniform distribution on $[a,b]$. These scenarios represent both symmetric and asymmetric contamination. In symmetric contamination, the outliers are symmetrically distributed around the true population mean, and therefore may have a relatively smaller impact on the estimate of the mean parameter. The detailed data-generating mechanisms for the three types of outliers follow Appendix B of \cite{zhao2023-rfpca}.

To compare the performance of different methods, we report the relative difference between the true covariance matrix$\bSig = \bSig_r \otimes \bSig_c$ and the estimated one $\bwSig$, measured as $| \bSig - \bwSig |_F / | \bSig |_F$, where $|\cdot|_F$ denotes the Frobenius norm. For HRFPCA, $\bwSig = \bwSig_r \otimes \bwSig_c$, where $\bwSig_c$ and $\bwSig_r$ are obtained from the reweighting step of HRFPCA in \refe{eqn:Sigc.est.h} and \refe{eqn:Sigr.est.h}. To reduce random variations, we report the average relative difference over 50 independent replications. The results, summarized in \reft{tab:breakdownpoints1} and \reft{tab:breakdownpoints2}, illustrate the effect of varying outlier proportions $p$ (10\%–50\%) under the four contamination scenarios.
\begin{table}[htbp]
	\centering
	\caption{Relative difference between the estimated and true covariance for multiple $p$'s in five cases (i.e., three kinds of outliers (PC, OC, and PC+OC) in symmetric and asymmetric cases, $\bU(-100,100)$ and $\bU(100, 110)$, respectively).\label{tab:breakdownpoints1}}
	\resizebox{\linewidth}{!}{
		\begin{tabular}{llllllllllll} 
			\toprule
			\multirow{2}{*}{Type} & \multirow{2}{*}{Method} & \multicolumn{10}{l}{Proportion $p$}   \\
			\cmidrule(r){3-12}
			&   & 10\% & 30\% & 40\% & 49\% & 50\% &  10\% & 30\% & 40\% & 49\% & 50\%  \\
			\midrule
			& & \multicolumn{5}{l}{Sit-\Rmn1: $\bU(-100,100)$}   & \multicolumn{5}{l}{Sit-\Rmn3: $\bU(100,110)$}   \\ 
			\cmidrule(r){3-7} \cmidrule(r){8-12}
			\multirow{4}{*}{PC}          & HRFPCA                       & 0.1           & 0.1           & 0.2           & 0.2           & 0.4           & 0.1           & 0.1           & 0.1           & 0.2           & 20.1           \\
			& RFPCA                        & 1.3           & 82.4          & 204.9         & 323.4         & 336.5         & 63.7          & 877           & 1051.9        & 1105.8        & 1106.3         \\
			& TPCA                         & 0.6           & 3.2           & 6.3           & 12.7          & 14            & 0.9           & 892.4         & 1065.2        & 1109.2        & 1109.5         \\
			& FPCA                         & 99.3          & 289.4         & 386.3         & 472.8         & 481.8         & 443.7         & 1328.4        & 1770.1        & 2167.3        & 2211.3         \\
			\addlinespace
			\multirow{4}{*}{OC}          & HRFPCA                       & 0.1           & 0.1           & 0.1           & 0.2           & 1             & 0.1           & 0.1           & 0.1           & 0.2           & 12.3           \\
			& RFPCA                        & 0.2           & 0.4           & 0.5           & 0.6           & 0.6           & 5             & 130           & 194.7         & 248.4         & 253.4          \\
			& TPCA                         & 0.5           & 0.4           & 0.6           & 1             & 1             & 0.2           & 162.7         & 330           & 355.5         & 351.5          \\
			& FPCA                         & 11.8          & 34.6          & 45.8          & 56.1          & 57.1          & 183.6         & 405.3         & 499.4         & 579.5         & 586.6          \\
			\addlinespace
			\multirow{4}{*}{PC+OC}         & HRFPCA                       & 0.1           & 0.1           & 0.1           & 0.7           & 0.8           & 0.1           & 0.1           & 0.1           & 0.2           & 80.9           \\
			& RFPCA                        & 0.1           & 0.4           & 0.6           & 0.5           & 0.5           & 15.4          & 515.2         & 857.9         & 1256.5        & 1310.4         \\
			& TPCA                         & 1             & 2.4           & 3.9           & 2.8           & 3.1           & 0.7           & 647.4         & 2523.9        & 4534.3        & 4720           \\
			& FPCA                         & 335.4         & 1002.3        & 1333.3        & 1637.4        & 1667.4        & 1686.5        & 5031.9        & 6686.9        & 8212.3        & 8347.2         \\
			\bottomrule
		\end{tabular}
	}
\end{table}
\begin{table}[htbp]
	\centering
	\caption{Relative difference between the estimated and true covariance for multiple $p$'s in five cases (i.e., three kinds of outliers (PC, OC, and PC+OC) in symmetric and asymmetric cases $\bU(-10000,10000)$ and $\bU(10000, 11000)$, respectively).\label{tab:breakdownpoints2}}
	\resizebox{\linewidth}{!}{
		\begin{tabular}{llllllllllll} 
			\toprule
			\multirow{2}{*}{Type} & \multirow{2}{*}{Method} & \multicolumn{10}{l}{Proportion $p$}   \\
			\cmidrule(r){3-12}
			&   & 10\% & 30\% & 40\% & 49\% & 50\% &  10\% & 30\% & 40\% & 49\% & 50\%  \\
			\midrule
			& & \multicolumn{5}{l}{Sit-\Rmn2: $\bU (-10000,10000)$}   & \multicolumn{5}{l}{Sit-\Rmn4: $\bU (10000,11000)$}   \\ 
			\cmidrule(r){3-7} \cmidrule(r){8-12}
			\multirow{4}{*}{PC}          & HRFPCA                       & 0.1           & 0.1           & 0.2           & 0.2           & 1622.6        & 0.1           & 0.1           & 0.2           & 0.2           & 211240         \\
			& RFPCA                        & 1.3           & 654670        & 1950000       & 3161500       & 3293700       & 100.9         & 4328400       & 9013000       & 10819000      & 10894000       \\
			& TPCA                         & 1             & 3.5           & 6.6           & 13.3          & 14.7          & 1.3           & 561.3         & 4203.7        & 14357         & 1145400        \\
			& FPCA                         & 965990        & 2863200       & 3828900       & 4696400       & 4786600       & 4402800       & 13216000      & 17625000      & 21595000      & 22029000       \\
			\addlinespace
			\multirow{4}{*}{OC}          & HRFPCA                       & 0.1           & 0.1           & 0.1           & 0.2           & 5491.8        & 0.1           & 0.1           & 0.1           & 0.2           & 62619          \\
			& RFPCA                        & 0.2           & 0.4           & 0.5           & 0.6           & 0.6           & 4.4           & 181.7         & 335.8         & 620.6         & 674.4          \\
			& TPCA                         & 0.5           & 0.4           & 0.6           & 1             & 1.1           & 0.5           & 134.2         & 780.7         & 2223          & 2507.8         \\
			& FPCA                         & 115270        & 342620        & 455110        & 558050        & 568120        & 746320        & 2208400       & 2951100       & 3618700       & 3680800        \\
			\addlinespace
			\multirow{4}{*}{PC+OC}         & HRFPCA                       & 0.1           & 0.1           & 0.1           & 0.2           & 126610        & 0.1           & 0.1           & 0.1           & 0.2           & 1095200        \\
			& RFPCA                        & 0.1           & 0.4           & 0.7           & 1             & 1             & 14.1          & 598.6         & 1054.5        & 1633.1        & 1717.2         \\
			& TPCA                         & 1             & 2.4           & 3.9           & 6.8           & 7.4           & 1.1           & 669.2         & 4134.7        & 11572         & 12962          \\
			& FPCA                         & 3354600       & 10025000      & 13332000      & 16375000      & 16676000      & 17034000      & 50368000      & 66919000      & 82253000      & 83606000       \\
			\bottomrule
		\end{tabular}
	}
\end{table}

\begin{enumerate}[(i)]
	\item \emph{HRFPCA versus FPCA}: HRFPCA exhibits strong robustness, in contrast to normal-based FPCA, which is highly sensitive to outliers. The relative difference of FPCA increases substantially with the contamination proportion $p$.	
	\item \emph{HRFPCA versus TPCA and RFPCA}: All three methods demonstrate some degree of robustness, yet their performance varies considerably. HRFPCA is robust in all cases for $p \leq 49\%$, exhibiting substantially greater robustness than both RFPCA and TPCA. Moreover, TPCA outperforms RFPCA; for example, in Sit-\Rmn3 with PC outliers at $p = 10\%$, the relative difference for RFPCA reaches 63.7, whereas that for TPCA is only 0.9, indicating that RFPCA breaks down under these conditions while TPCA remains stable. Overall, the comparative robustness of the methods can be ranked as HRFPCA $>$ TPCA $>$ RFPCA.
\end{enumerate}

The robustness of an estimator is commonly assessed via its breakdown point. Under the settings of this experiment ($d_c = 4$, $d_r = 10$, $n = 1000$), if the vectorized data were modeled using a multivariate $t$ distribution, the breakdown point would be $1/(d_cd_r+\nu) < 1/(d_cd_r) = 2.5\%$, which is extremely low and would deteriorate further as the data dimensionality increases. As noted by \cite{zhao2023-rfpca}, although the breakdown points of TPCA and RFPCA have not been theoretically established, empirical evidence suggests that both exceed 2.5\%, highlighting a key motivation for developing matrix-based methods. According to \cite{Mayrhofer2025-mmcd}, the maximum theoretical breakdown point of HRFPCA is $\lfloor(n-d)/2\rfloor/n = 49.9\%$, and experimental results further confirm that the breakdown points of TPCA and RFPCA are substantially lower than that of HRFPCA.

\subsection{Outlier dection and explanation}\label{sec:outflag.syn.expr}
This experiment employs \emph{Data2} to evaluate the capability of HRFPCA, relative to FPCA, in detecting and explaining outliers. To simulate outlying data, 40 observations (i.e., $p=20\%$) were randomly selected and replaced with outliers originating from adding three types of outliers: PC outliers, OC outliers, and PC+OC outliers. The uniform distribution required for generating these three types of outliers was set to Sit-\Rmn3, i.e. $\bU(100, 110)$. For convenience, the modified dataset is denoted as \emph{Data2-O}.

\begin{figure}
	\centering
	\scalebox{0.7}[0.7]{\includegraphics*{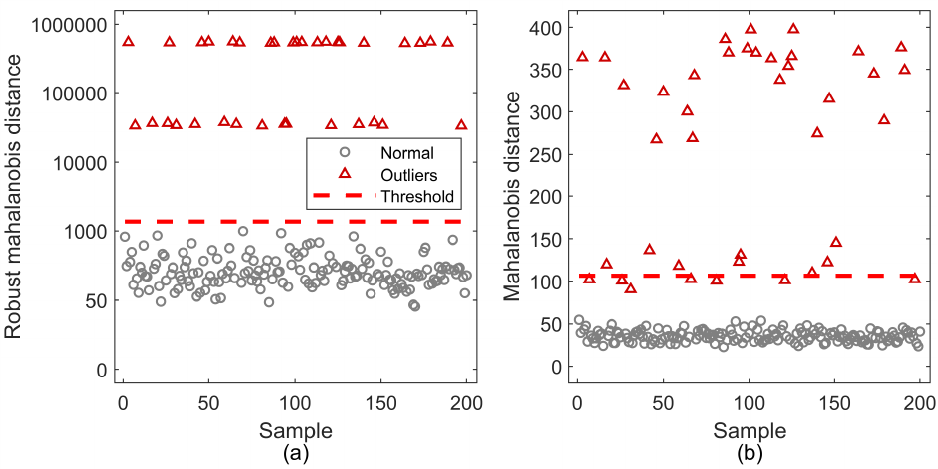}}
	\caption{Outlier dection based on (a) robust matrix mahalanobis distance by the MMCD estimators; (b) matrix mahalanobis distance.}
	\label{fig:OutlierDet}
\end{figure}

Stage 1 of HRFPCA was then applied to \emph{Data2-O}, yielding the MMCD estimates $\bwM^H$, $\bwSig_c^H$, and $\bwSig_r^H$. In parallel, the `\textit{flip-flop}' algorithm was executed on \emph{Data2-O} to obtain the ML estimates in \refe{eqn:M.est}-\refe{eqn:Sigr.est}, namely $\bwM$, $\bwSig_c$, and $\bwSig_r$. To detect outliers, both sets of estimates were substituted into \refe{eqn:out.dect}. \reff{fig:OutlierDet} presents the squared matrix Mahalanobis distances derived from the two estimation procedures, along with the corresponding 97.5\% quantile critical value ($\chi^{2}_{q_cq_r;0.975}$) of the chi-square distribution. Evidently, regardless of outlier type, the robust matrix Mahalanobis distance based on the MMCD estimates successfully identifies all outlier points. In contrast, the squared matrix Mahalanobis distance obtained by ML estimates under the matrix normal distribution fails to identify all outliers.

\begin{figure}
	\centering
	\scalebox{0.75}[0.75]{\includegraphics*{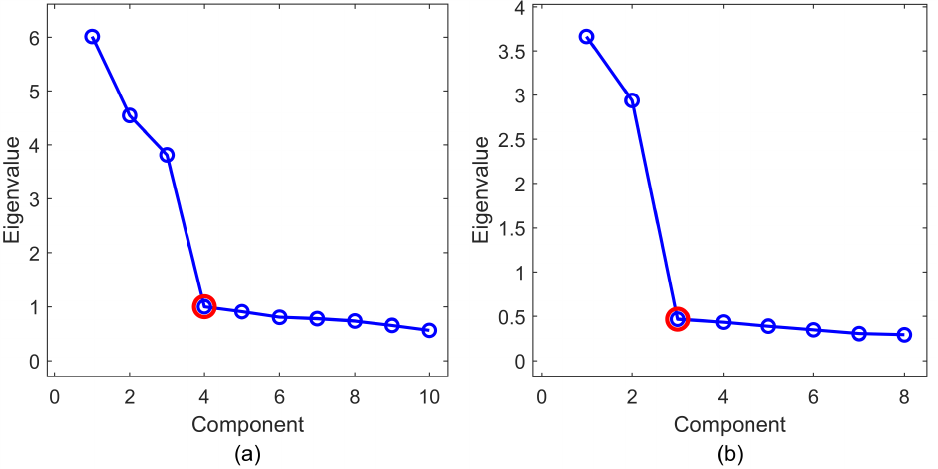}}
	\caption{The scree plot of robust covariance matrices $\bwSig_c^H$ and $\bwSig_r^H$ based on the MMCD estimates in (a) columns; (b) rows.}
	\label{fig:hrfpca-scree}
\end{figure}

To determine the number of row and column PCs for HRFPCA and FPCA, eigenvalue decompositions were performed on $\bwSig_c^H$, $\bwSig_r^H$, $\bwSig_c$, and $\bwSig_r$, respectively. It can be observed that the scree plots for the robust covariance matrices $\bwSig_c^H$ and $\bwSig_r^H$, suggesting PC numbers $(4,3)$ for HRFPCA in \reff{fig:hrfpca-scree}. The scree plots (not shown for brevity) for $\bSig_c$ and $\bSig_r$ recommend PC numbers $(3,4)$ for FPCA.

\begin{figure}
	\centering
	\scalebox{0.7}[0.7]{\includegraphics*{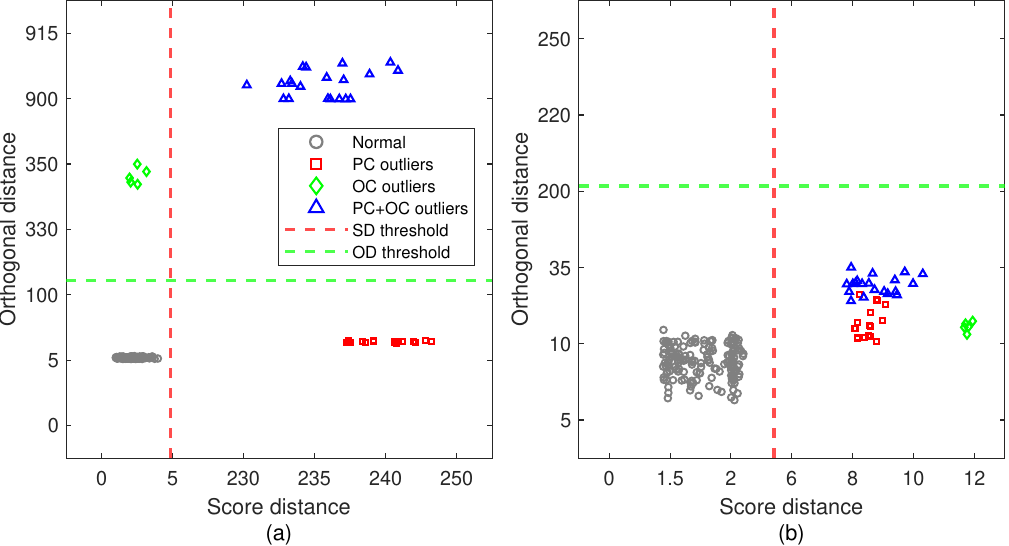}}
	\caption{The SODA plot obtained by (a) HRFPCA; (b) FPCA.}
	\label{fig:DDplot}
\end{figure}

Using the PC numbers obtained above, HRFPCA and FPCA were executed to obtain the principal component score matrices $\bZ^H$ and $\bZ$, respectively. Score distances (SD) and orthogonal distances (OD) were then calculated according to \refe{eqn:SD} and \refe{eqn:OD}. The SODA plot introduced in \refs{sec:diag.plot} was employed to identify the three types of outliers. \reff{fig:DDplot} displays the SODA plots for HRFPCA and FPCA. It is evident that the SODA plot based on HRFPCA accurately identifies all three types of outliers. In contrast, the SODA plot derived from FPCA misclassifies the OC and PC+OC outliers as PC outliers. This indicates that the subspace estimated by FPCA is influenced by bad leverage points, causing the principal component score matrix $\bZ$ to be contaminated with excessive outlier information, which inflates the score distances. This observation aligns with the findings reported in \cite{Hubert2005-robpca}.

\begin{figure}
	\centering
	\scalebox{0.35}[0.4]{\includegraphics*{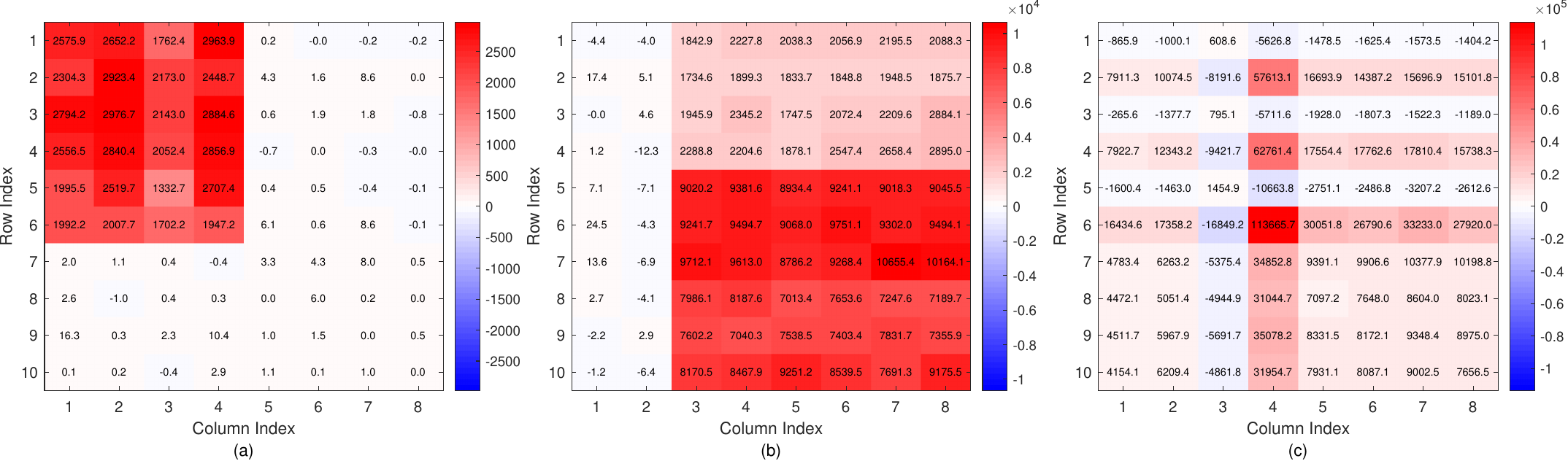}}
	\caption{The heat maps of Shapley values for the first observation from (a) PC outliers; (b) OC outliers; (c) PC+OC outliers.}
	\label{fig:Shapley}
\end{figure}

To further investigate the mechanisms underlying the outliers, the first observation from each of the three outlier types in \emph{Data2-O}—labeled PC, OC, and PC+OC—was extracted. \reff{fig:Shapley} displays heat maps of the Shapley values for each observation, computed according to equation \refe{eqn:cellwise} in \refs{sec:out.dect}. The results clearly indicate that each type of outlier induces a distinct pattern of alterations in the original data, thereby highlighting the different ways in which these outliers exert their influence.

\section{Applications}\label{sec:real.expr}
This section conducts a series of experiments on real-world datasets, including face datasets and multivariate time series datasets, to compare the classification performance of the proposed HRFPCA with three related matrix-based PCA methods (FPCA, TPCA, and RFPCA). 

\subsection{Performance of HRFPCA on face datasets}\label{sec:face.real.expr}
This section primarily conducts experiments on the following two real face datasets to evaluate the classification performance of HRFPCA.

\textbullet\ XM2VTS. This dataset comprises images of 295 individuals collected across four sessions spanning four months \citep{Messer1999XM2VTSDBTE}. Each session contains two images per person, resulting in a total of 8 images per individual. Each image has dimensions of $51\times55$. 

\textbullet\ PIE . This dataset includes 41,368 facial images of 68 individuals, captured under 13 different poses, 43 lighting conditions, and 4 facial expressions \citep{Gross2010-pie}. In our experiments, we adopt the same subset as in \cite{zhao2015-rlda-2stage}, consisting of 43 frontal images per individual under varying lighting conditions, each of size $64\times64$. 

\begin{figure}[htbp]
	\centering
	\subfigure[XM2VTS]{\scalebox{0.5}[0.5]{\includegraphics*{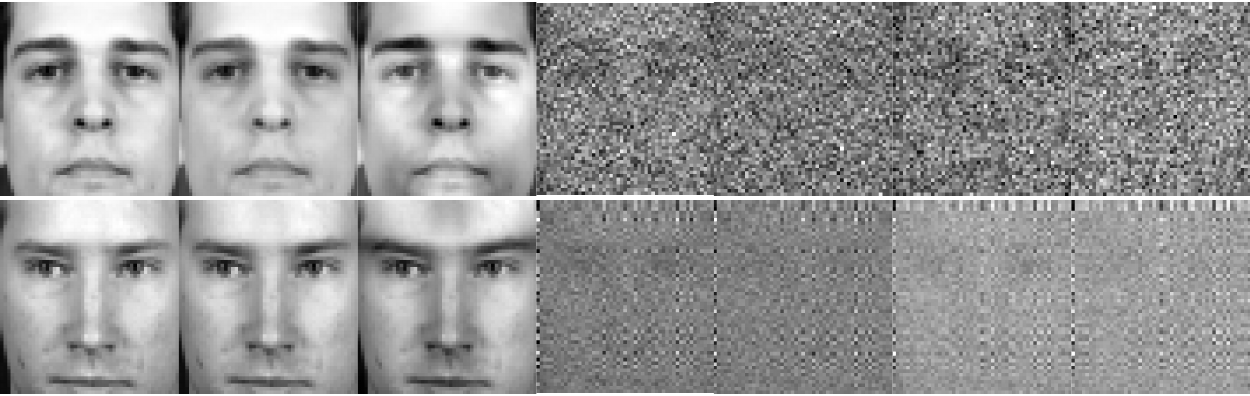}}}  \\
	\subfigure[PIE]{\scalebox{0.5}[0.5]{\includegraphics*{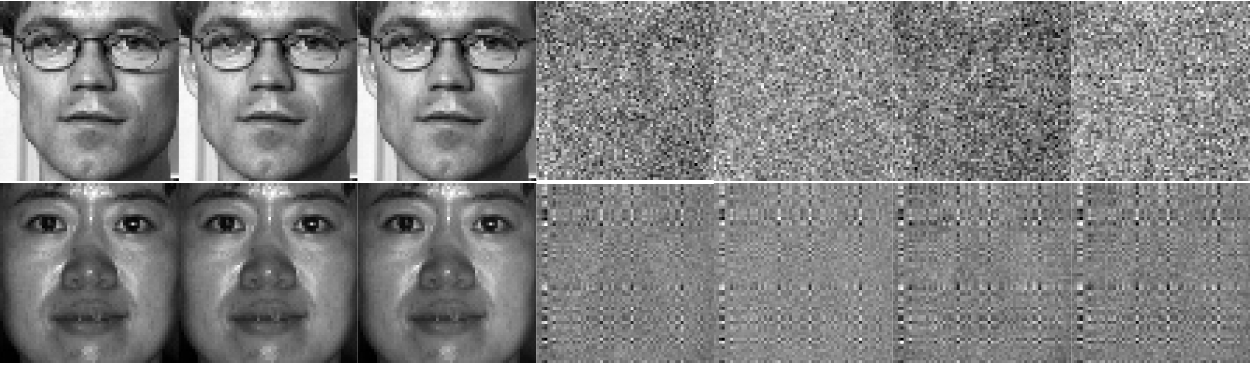}}}		
	\caption{Example images from the two face data sets and tree types of outlying images.}
	\label{fig:images}
\end{figure}

For both datasets, the images are partitioned into training and test sets. Specifically, $\gamma$ images are randomly selected from each individual to form a training set of size $n_{tr}$, while the remaining $n_{ts}$ images comprise the test set. For XM2VTS, $\gamma = 4$, and for PIE, $\gamma = 6$. To evaluate robustness, $n_{tr}p$ outlying images are replaced into the training set. Two experimental scenarios are considered:
\begin{enumerate}[(i)]
	\item \emph{Scenario 1}. Two types of outlying images are considered: (1) Type \Rmn1: images occluded by dummy patches; (2) Type \Rmn2: dummy images. The dummy images and patches consist of black and white dots, obtained by randomly setting their pixel values to be 0 or 1 with a probability of 0.5. For Type \Rmn1, one image from each person in the training set is randomly selected and occluded at a random location. The patch sizes range from $30 \times 30$ to $50 \times 50$. Three cases are considered: (1) Case \Rmn1: Type \Rmn1, $p=1/\gamma$; (2) Case \Rmn2: Type \Rmn2, $p=10\%$; (3) Case \Rmn3: Type \Rmn2, $p=20\%$. Example images can be found in \cite{zhao2023-rfpca}.
	\item \emph{Scenario 2}. Four cases involving three types of outlying images (PC, OC, and PC+OC) are considered: (1) Case \Rmn1: $p=10\%, \bU(-10,10)$; (2) Case \Rmn2: $p=20\%, \bU(-10,10)$; (3) Case \Rmn3: $p=10\%, \bU(0,10)$; and (4) Case \Rmn4: $p=20\%, \bU(0,10)$, resulting in 12 outlying cases in total. \reff{fig:images} shows example images from the XM2VTS and PIE datasets under different outliers. From left to right, the images correspond to the original image, 10\% PC outliers, 20\% PC outliers, 10\% OC outliers, 20\% OC outliers, 10\% PC+OC outliers, and 20\% PC+OC outliers. The two rows represent outliers generated from $\bU(-10,10)$ and $\bU(0,10)$, respectively. It is evident that PC outliers primarily distort key facial regions such as the nose, mouth, eyes, and forehead, without drastically altering the overall facial structure. In contrast, OC and PC+OC outliers nearly completely obscure the facial image, leaving only a blurred outline, which significantly impairs face recognition.
\end{enumerate}	

In this experiment, we compare the classification performance of HRFPCA with RFPCA, TPCA, and FPCA. For all four methods, the scaled compression in the $(q_c,q_r)$-dimensional space is represented as $\bZ = \bLmd_c^{-1/2} \bU_c' \bX \bU_r \bLmd_r^{-1/2}$. For all methods, this experiment tries all possible dimensions (i.e., all possible values of $(q_c,q_r)$) for the compressed representation and uses a nearest neighbor classifier in the reduced-dimensional space to obtain the classification error rates. To assess whether the performance differences between HRFPCA and the other three methods are statistically significant, we conduct pairwise comparisons using the Wilcoxon signed-rank test between HRFPCA and each competing method.

\reft{tab:face.class} and \reft{tab:face.class.2} report the lowest average classification error rates, their standard deviations, and the corresponding latent dimensions achieved by each method across 20 random splits for the XM2VTS and PIE face datasets under the two experimental scenarios, respectively. The best-performing results are highlighted in bold. The key findings can be summarized as follows:
\begin{enumerate}[(i)]
	\item HRFPCA achieves the best classification performance across all considered cases. RFPCA and TPCA exhibit comparable results, and both generally outperform FPCA.
	\item HRFPCA, RFPCA, and TPCA demonstrate strong robustness, as their recognition rates and corresponding latent dimensions remain relatively stable regardless of the presence of outliers in the dataset. In contrast, FPCA exhibits poor robustness: its performance frequently deteriorates in the presence of outliers and typically requires a greater number of features to achieve optimal recognition performance.
	\item Compared with OC and PC+OC outliers, PC outliers exert a milder effect on the recognition rates of all methods and may even yield slight performance gains in certain cases. However, recognition performance deteriorates as the proportion of PC outliers increases.
	\item HRFPCA demonstrates lower sensitivity to the type of outlier than the other three methods, consistently yielding strong performance regardless of the outlier type. For instance, on the PIE dataset, even under OC outliers in Case \Rmn3 and Case \Rmn4, HRFPCA achieves error rates of 9.4 and 9.5, respectively—both lower than FPCA’s error rate of 9.7 on the clean data.
\end{enumerate}
\begin{table}[htbp]
	\centering
	\caption{The lowest average error rates (mean ± standard deviation) and corresponding reduced dimensions for different methods under \emph{Scenario 1}. The best method is highlighted in bold. Using the Wilcoxon signed-rank test at a 95$\%$ confidence level, • indicates HRFPCA is significantly better than the corresponding method, ◦ indicates the difference between HRFPCA and the corresponding method is not significant. \label{tab:face.class}}
	\resizebox{\linewidth}{!}{
	\begin{tabular}{llllll} 
		\toprule
		\multirow{2}{*}{Dataset} & \multirow{2}{*}{Method} & \multirow{2}{*}{Without outliers}  & \multicolumn{3}{l}{With outliers}   \\
		\cmidrule(r){4-6}
		&    &   & Case \Rmn1  & Case \Rmn2 & Case \Rmn3  \\ 
		\midrule
		\multirow{4}{*}{XM2VTS}   & HRFPCA   & \textbf{10.6 ±2.1(16,10)}        & \textbf{10.6±2.4(15,9)}          & \textbf{11.3\textbf{±}2.4(16,10)} & \textbf{12.1\textbf{±}2.2(17,9)}  \\
		& RFPCA                        & 11.1±1.7(17,7)◦                 & 11.3±2.1(12,7)•                  & 11.6±2.2(13,7)◦                   & 12.7±2.0(13,8)•                   \\
		& TPCA                         & 11.3
		±   
		1.9(12,7)•           & 11.2±1.9(12,7)◦                  & 11.8±2.1(21,5)◦                   & 12.5±2.4(21,7)•                   \\
		& FPCA                         & 11.4
		±   
		1.9(12,8)•           & 11.6±1.8(19,8)•                  & 13.5±2.0(26,10)•                  & 14.2±2.1(25,13)•                  \\ 
		\addlinespace
		\multirow{4}{*}{PIE}          & HRFPCA                       & \textbf{9.1\textbf{±}5.1(25,20)} & \textbf{9.1\textbf{±}5.1(25,20)} & \textbf{9.4\textbf{±}5.3(26,21)}  & \textbf{9.8\textbf{±}5.2(26,18)}  \\
		& RFPCA                        & 9.4±4.8(27,15)◦                  & 9.3±4.7(27,15)◦                  & 9.7±4.9(27,16)◦                   & 10.1±5.1(28,12)•                  \\
		& TPCA                         & 9.4±4.7(27,15)◦                  & 9.4±4.7(26,16)◦                  & 9.7±4.8(27,18)◦                   & 10.6±4.7(28,18)•                  \\
		& FPCA                         & 9.7±4.9(29,12)•                  & 10.0±5.2(54,13)•                 & 11.2±5.6(64,21)•                  & 12.1±5.8(64,35)•                  \\
		\bottomrule
	\end{tabular}
	}
\end{table}
\begin{table}[htbp]
	\centering
	\caption{The lowest average error rates (mean ± standard deviation) and corresponding reduced dimensions for different methods under \emph{Scenario 2}. The best method is highlighted in bold. Using the Wilcoxon signed-rank test at a 95$\%$ confidence level, • indicates HRFPCA is significantly better than the corresponding method, ◦ indicates the difference between HRFPCA and the corresponding method is not significant. \label{tab:face.class.2}}
\resizebox{\linewidth}{!}{
\begin{tabular}{lllllll} 
			\toprule
			\multirow{3}{*}{Dataset} & \multirow{3}{*}{Type}  & \multirow{3}{*}{Method}  & \multicolumn{4}{l}{With outliers}                                          \\
			\cmidrule(r){4-7}
			&                        &                          & \multicolumn{2}{l}{$\bU(-10,10)$}       & \multicolumn{2}{l}{$\bU(0,10)$}          \\
			\cmidrule(r){4-5} \cmidrule(r){6-7}
			&                        &                         & Case \Rmn1           & Case \Rmn2          & Case \Rmn3         & Case \Rmn4           \\ 
			\midrule
			\multirow{12}{*}{XM2VTS} & \multirow{4}{*}{PC}    & HRFPCA                  & \textbf{10.6±2.0(16,10)}  & \textbf{10.8±2.1(18,10)}  & \textbf{10.9±2.1(18,7)}   & \textbf{11.2±2.3(18,10)}   \\
			&                        & RFPCA                   & 11.3±2.1(18,6)•  & 11.5±1.8(19,7)◦  & 11.7±2.0(18,6)◦  & 12.2±2.2(20,7)•   \\
			&                        & TPCA                    & 11.2±1.9(12,7)◦  & 11.6±1.8(14,8)◦  & 11.6±2.0(12,8)◦  & 12.1±2.0(17,7)•   \\
			&                        & FPCA                    & 11.6±1.9(19,8)•  & 12.2±2.5(25,14)• & 12.1±2.9(23,12)• & 12.6±3.1(25,18)•  \\
			\addlinespace
			& \multirow{4}{*}{OC}    & HRFPCA                  & \textbf{10.8±2.2(18,7)}   & \textbf{11.2±2.1(18,7)}   & \textbf{11.4±2.1(17,9)}   & \textbf{11.9±2.4(19,10)}   \\
			&                        & RFPCA                  & 12.3±2.1(18,7)•  & 12.8±1.8(19,7)•  & 13.1±2.0(21,5)•  & 13.6±2.2(23,5)•   \\
			&                        & TPCA                    & 12.2±2.1(12,8)•  & 12.9±2.2(14,10)• & 13.1±2.0(12,13)• & 13.5±2.3(18,10)•  \\
			&                        & FPCA                     & 12.4±1.9(19,15)• & 13.2±2.5(25,18)• & 13.7±2.9(27,15)• & 14.9±3.1(35,20)•  \\
			\addlinespace
			& \multirow{4}{*}{PC+OC} & HRFPCA                  & \textbf{11.4±2.3(18,12)}  & \textbf{11.9±2.5(19,13)}  & \textbf{12.3±2.5(19,10)}  & \textbf{12.9±2.3(19,15)}   \\
			&                        & RFPCA                   & 12.6±2.2(19,7)•  & 13.7±2.3(21,7)•  & 13.9±2.0(19,8)•  & 14.8±2.2(23,8)•   \\
			&                        & TPCA                    & 12.7±2.2(14,12)• & 13.4±2.3(17,13)• & 13.5±2.0(15,12)• & 14.5±2.4(19,15)•  \\
			&                        & FPCA                   & 13.2±3.1(31,20)• & 14.1±3.3(42,22)• & 14.9±2.5(42,22)• & 16.3±2.3(55,24)•  \\ 
			\hline
			\multirow{12}{*}{PIE}    & \multirow{4}{*}{PC}    & HRFPCA                   & \textbf{9.0±5.1(26,20)}   & \textbf{9.2±5.0(26,21)}   & \textbf{9.1±5.1(26,20)}   & \textbf{9.3±5.3(27,21)}    \\
			&                        & RFPCA                   & 9.6±4.7(27,15)◦  & 9.9±4.8(28,16)◦  & 9.7±4.8(28,15)◦  & 10.1±5.3(29,16)•  \\
			&                        & TPCA                     & 9.7±4.1(26,16)◦  & 10.0±4.0(28,18)◦ & 9.9±4.1(28,16)◦  & 10.2±4.5(29,17)•  \\
			&                        & FPCA                     & 9.9±5.2(32,13)•  & 10.2±5.1(36,15)• & 10.2±5.1(35,12)• & 10.5±5.3(41,13)•  \\
			\addlinespace
			& \multirow{4}{*}{OC}    & HRFPCA                  & \textbf{9.2±5.2(27,20)}   & \textbf{9.4±5.5(29,21)}   & \textbf{9.5±5.3(27,21)}   & \textbf{9.8±5.4(31,21)}    \\
			&                        & RFPCA                  & 9.8±4.9(28,16)◦  & 10.2±5.2(31,18)• & 10.5±5.1(27,18)• & 11.1±5.0(32,17)•  \\
			&                        & TPCA                     & 9.9±4.3(28,18)◦  & 10.4±4.5(32,16)• & 10.7±4.5(30,15)• & 11.3±4.7(32,21)•  \\
			&                        & FPCA                     & 11.2±5.3(38,18)• & 11.8±5.0(44,25)• & 12.1±5.2(43,27)• & 12.8±5.1(51,30)•  \\
			\addlinespace
			& \multirow{4}{*}{PC+OC} & HRFPCA                   & \textbf{9.3±5.2(28,21)}   & \textbf{9.7±5.1(31,23)}   & \textbf{9.8±5.3(29,22)}   & \textbf{10.2±5.4(35,20)}   \\
			&                        & RFPCA                   & 10.1±5.1(31,17)• & 10.7±5.3(35,17)• & 10.6±5.0(33,18)• & 11.3±5.1(39,17)•  \\
			&                        & TPCA                     & 10.3±4.5(32,17)• & 11.1±4.7(34,16)• & 11.4±4.8(36,17)• & 11.9±4.8(38,21)•  \\
			&                        & FPCA                    & 11.9±5.3(43,27)• & 12.5±5.4(50,28)• & 12.6±5.3(52,31)• & 13.5±5.5(64,33)•\\
			\bottomrule
	\end{tabular}}
\end{table}

\subsection{Performance of HRFPCA on matrix-variate time series}\label{sec:timeseries}
Multivariate time series (MTS) represent a common type of matrix-valued data, structured as observation matrices where the two dimensions correspond to variables and time points, respectively. This structure contrasts sharply with the image data analyzed in \refs{sec:face.real.expr}, where both dimensions of the data matrix represent pixels. Given this fundamental difference, matrix-variate methods that assume a separable covariance structure are intuitively more suitable for MTS. Therefore, this section evaluates the classification performance of such methods on two publicly available, real-world MTS datasets.


\textbullet\ AUSLAN. This dataset contains 2565 observations captured by 22 sensors (i.e., 22 variables) on a CyberGlove from a native AUSLAN speaker \citep{Kadous2002-auslan}. The dataset contains 95 gestures (i.e., 95 classes), with 27 observations per gesture. For the experiments in this section, a subset of 25 gestures is used (thus totaling 675 observations). The time length of these 675 samples ranges from 47 to 95; they are uniformly truncated to the shortest length of 47, resulting in observations of size $47 \times 22$. 

\textbullet\ ECG. This dataset contains 200 multivariate time series observations, each collected by two electrodes (i.e., 2 variables) during a single heartbeat \citep{Olszewski2001-ECG200}. The dataset contains two classes: normal and abnormal, with 133 and 67 observations, respectively. The time length of the 200 samples ranges from 39 to 152; they are uniformly truncated to the shortest length of 39, resulting in observations of size $39 \times 22$. Due to the large numerical range of the ECG dataset, a normalization transformation is applied as a preprocessing step to eliminate scale effects. 

From each class, a proportion $\gamma = 4/5$ of observations are randomly selected as the training set, and the remaining observations form the test set. To test the robustness of the four PCA-type methods, $n_{tr}p$ outliers matrices are replaced within the training set. The generation of outliers primarily considers two schemes. \emph{Scenario 1:} All elements of the outliers matrix are drawn from a given uniform distribution $\bU$; \emph{Scenario 2:} Three types of outliers matrices (PC, OC, and PC+OC) are generated separately, following the method described in \cite{zhao2023-rfpca}, which also requires specifying a uniform distribution $\bU$. This experiment considers four cases under both symmetric and asymmetric contamination with two proportions: (1) Case \Rmn1: $p=10\%, \bU(-10,10)$; (2) Case \Rmn2: $p=20\%, \bU(-10,10)$; (3) Case \Rmn3: $p=10\%, \bU(0,10)$; and (4) Case \Rmn4: $p=20\%, \bU(0,10)$.

To measure the classification error rate of different methods, similar to the face data, a simple 1-nearest neighbor classifier is used in the reduced-dimensional space to obtain the classification error rates. \reft{tab:mts.class} and \reft{tab:mts.class.2} summarize the results over 20 random splittings for the AUSLAN and ECG datasets under the two scenarios, respectively. The symbol `--' indicates that the method failed to run, thus the error rate is unavailable. The main observations are largely consistent with those in \refs{sec:face.real.expr}, but some additional conclusions can be drawn:
\begin{enumerate}[(i)]
	\item HRFPCA outperforms all competing methods in every case except on the original AUSLAN dataset without outliers, where FPCA achieves the best performance. However, FPCA exhibits severe performance degradation once outliers are introduced. Under Case \Rmn3 and Case \Rmn4, the classification improvements of RFPCA and TPCA over FPCA are not statistically significant, whereas HRFPCA delivers a substantial and statistically significant performance gain.
	\item Whether the AUSLAN dataset contains outliers or not, the TPCA implementation based on the MATLAB program \citep{zhao2023-rfpca} and the TPCA implementation in the R package \texttt{MixMatrix} fail to run due to numerical issues.
	\item Compared with image data, HRFPCA exhibits a more pronounced performance improvement over competing methods on MTS.
\end{enumerate}
\begin{table}[htbp]
	\centering
	\caption{The lowest average error rate (mean ± standard deviation) and its corresponding reduced dimensionality for different methods under \emph{Scenario 1}. The best method is highlighted in bold. Using the Wilcoxon signed-rank test at a 95\% confidence level, • indicates that HRFPCA is significantly better than the corresponding method, while ◦ indicates that the difference between HRFPCA and the corresponding method is not significant.\label{tab:mts.class}}
	\resizebox{\linewidth}{!}{
		\begin{tabular}{lllllll} 
			\toprule
			\multirow{3}{*}{Dataset} & \multirow{3}{*}{Method} & \multirow{3}{*}{Without outliers} & \multicolumn{4}{l}{With outliers}   \\
			\cmidrule(r){4-7} 
			& & & \multicolumn{2}{l}{$\bU(-10,10)$} & \multicolumn{2}{l}{$\bU(0,10)$} \\
			\cmidrule(r){4-5} \cmidrule(r){6-7}
			&    &   & Case \Rmn1   & Case \Rmn2   & Case \Rmn3   & Case \Rmn4     \\ 
			\midrule
			\multirow{4}{*}{AUSLAN}       & HRFPCA                       & 4.0±2.1(2,11)                   & \textbf{4.2±2.3(2,11)}  & \textbf{4.8±2.4(2,12)}  & \textbf{4.2±2.4(2,11)}  & \textbf{5.2±2.6(2,11)}   \\
			& RFPCA                        & 4.8±1.7(1,12)◦                  & 5.2±2.1(1,13)◦          & 6.4±2.2(1,13)•          & 6.4±2.2(1,13)•          & 6.4±2.3(1,13)•           \\
			& TPCA         & --       & --    & --   & --   & --      \\
			& FPCA                         & \textbf{2.4±1.3(1,14)◦}         & 7.2±3.2(1,20)•          & 11.2±3.1(2,13)•         & 6.5±2.1(3,16)•          & 6.6±2.1(3,22)•           \\ 
			\addlinespace
			\multirow{4}{*}{ECG}          & HRFPCA                       & \textbf{11.9±5.3(11,1)}         & \textbf{11.6±5.1(12,2)} & \textbf{12.5±5.9(12,2)} & \textbf{12.5±4.2(13,2)} & \textbf{13.2±4.2(18,2)}  \\
			& RFPCA                        & 12.5±4.8(11,1)◦                 & 12.5±4.7(10,1)◦         & 13.2±4.8(11,2)◦         & 12.7±5.1(14,2)◦         & 15.6±4.9(15,2)•          \\
			& TPCA                         & 12.1±3.7(11,1)◦                 & 14.5±4.5(25,1)•         & 15.0±4.8(25,2)•         & 15.5±4.7(33,2)•         & 16.5±5.2(34,1)•          \\
			& FPCA                         & 14.0±4.9(8,1)◦                  & 17.5±5.2(25,1)•         & 17.8±5.4(25,1)•         & 16.5±5.1(29,2)•         & 16.8±5.4(27,2)•          \\
			\bottomrule
		\end{tabular}
	}
\end{table}
\begin{table}[htbp]
	\centering
	\caption{The lowest average error rate (mean ± standard deviation) and its corresponding reduced dimensionality for different methods under \emph{Scenario 2}. The best method is highlighted in bold. Using the Wilcoxon signed-rank test at a 95\% confidence level, • indicates that HRFPCA is significantly better than the corresponding method, while ◦ indicates that the difference between HRFPCA and the corresponding method is not significant.\label{tab:mts.class.2}}
	\resizebox{\linewidth}{!}{
		\begin{tabular}{lllllll} 
			\toprule
			\multirow{3}{*}{Dataset} & \multirow{3}{*}{Type}  & \multirow{3}{*}{Method}  & \multicolumn{4}{l}{With outliers}                                      \\
			\cmidrule(r){4-7}
			&                        &                  & \multicolumn{2}{l}{$\bU(-10,10)$}     & \multicolumn{2}{l}{$\bU(0,10)$}        \\
			\cmidrule(r){4-5} \cmidrule(r){6-7}
			&                        &                  & Case \Rmn1          & Case \Rmn2         & Case \Rmn3        & Case \Rmn4    \\ 
			\midrule
			\multirow{12}{*}{AUSLAN} & \multirow{4}{*}{PC}    & HRFPCA        & \textbf{4.2±2.0(2,11)}   & \textbf{4.5±2.1(2,11)}   & \textbf{4.3±2.1(2,11)}   & \textbf{4.8±2.3(2,12)}    \\
			&                        & RFPCA                  & 5.2±2.1(1,13)◦  & 5.6±1.8(1,14)◦  & 5.5±2.0(1,12)◦  & 6.2±2.2(1,14)•   \\
			&                        & TPCA                   & —               & —               & —               & —                \\
			&                        & FPCA                   & 7.1±1.9(1,15)•  & 8.0±2.5(2,14)•  & 7.4±2.9(3,12)•  & 9.6±3.1(2,18)•   \\
			\addlinespace
			& \multirow{4}{*}{OC}    & HRFPCA                 & \textbf{4.3±2.1(2,11)}   & \textbf{4.8±2.3(3,11)}   & \textbf{4.5±2.4(2,11)}   & \textbf{5.0±2.6(3,12)}    \\
			&                        & RFPCA                  & 5.4±2.4(1,12)◦  & 6.2±3.2(1,13)•  & 6.7±3.1(1,10)•  & 8.3±3.3(1,13)•   \\
			&                        & TPCA                   & —               & —               & —               & —                \\
			&                        & FPCA                   & 9.6±1.8(2,20)•  & 10.4±2.1(1,22)• & 10.7±2.3(2,22)• & 11.2±2.2(2,22)•  \\
			\addlinespace
			& \multirow{4}{*}{PC+OC} & HRFPCA                 & \textbf{4.2±2.3(2,12)}   & \textbf{4.5±2.5(3,11)}   & \textbf{4.1±2.5(2,11)}   & \textbf{5.2±2.4(3,12)}    \\
			&                        & RFPCA                  & 5.2±2.2(1,13)◦  & 6.4±2.3(2,12)•  & 7.1±2.0(2,11)•  & 8.8±2.2(3,11)•   \\
			&                        & TPCA                   & —               & —               & —               & —                \\
			&                        & FPCA                   & 10.4±3.1(1,20)• & 11.2±3.3(2,18)• & 11.5±2.5(2,21)• & 12.0±2.3(3,22)•  \\ 
			\cmidrule{2-7}
			\multirow{12}{*}{ECG}    & \multirow{4}{*}{PC}    & HRFPCA         & \textbf{11.5±5.1(11,2)}  & \textbf{12.5±5.0(12,2)}  & \textbf{12.0±5.1(12,1)}  & \textbf{12.6±5.3(12,2)}   \\
			&                        & RFPCA                  & 12.8±4.7(11,1)◦ & 13.4±4.8(13,1)◦ & 13.1±4.8(12,2)◦ & 14.3±5.0(14,2)◦  \\
			&                        & TPCA                   & 13.4±4.1(11,1)◦ & 14.0±4.0(13,2)◦ & 14.4±4.1(14,2)◦ & 15.0±4.5(17,1)•  \\
			&                        & FPCA                   & 15.2±5.2(11,1)• & 15.8±5.1(14,2)• & 17.4±5.1(12,1)• & 18.6±4.9(17,2)•  \\
			\addlinespace
			& \multirow{4}{*}{OC}    & HRFPCA                 & \textbf{11.4±5.1(12,1)}  & \textbf{12.3±5.4(13,2)}  & \textbf{12.2±5.3(12,2)}  & \textbf{12.5±5.4(15,2)}   \\
			&                        & RFPCA                  & 13.2±4.9(11,1)◦ & 14.2±5.2(11,3)◦ & 14.0±5.1(12,2)◦ & 15.8±5.0(13,2)•  \\
			&                        & TPCA                   & 14.8±4.3(19,1)• & 16.1±4.1(22,2)• & 16.5±4.5(25,1)• & 17.8±4.7(27,2)•  \\
			&                        & FPCA                   & 16.9±5.1(25,1)• & 17.4±5.3(22,1)• & 17.6±5.0(19,2)• & 19.2±5.1(25,2)•  \\
			\addlinespace
			& \multirow{4}{*}{PC+OC} & HRFPCA                 & \textbf{11.6±5.2(12,2)}  & \textbf{12.4±5.1(15,2)}  & \textbf{12.6±5.1(15,2)}  & \textbf{13.0±5.4(18,3)}   \\
			&                        & RFPCA                  & 12.8±5.1(13,1)◦ & 14.1±5.3(15,2)◦ & 14.0±5.0(15,1)◦ & 15.8±4.9(17,2)•  \\
			&                        & TPCA                   & 15.2±4.5(25,1)• & 15.8±4.7(30,2)• & 16.7±4.8(33,1)• & 19.4±5.1(36,1)•  \\
			&                        & FPCA                   & 17.4±5.3(19,2)• & 20.1±5.1(22,1)• & 18.0±5.4(25,2)• & 19.6±5.3(29,3)•  \\
			\bottomrule
		\end{tabular}
	}
\end{table}

\section{Conclusion and future works}\label{sec:discussion} 
This paper introduces a novel, highly robust factored PCA for matrix-valued data, termed HRFPCA. The key innovation lies in replacing the covariance estimates in FPCA with the robust MMCD estimates. As a result, HRFPCA combines the dimension-reduction capability of FPCA for matrix-valued data with the high robustness of the MMCD estimates, achieving a theoretical breakdown point approaching 50\%. Additionally, we propose a SODA plot based on HRFPCA, which effectively displays and classifies different types of outliers. Extensive simulations and real-data experiments demonstrate that HRFPCA outperforms existing matrix-based PCA methods.

For vector-valued data, robust factor analysis based on the MCD estimates was proposed by \cite{pison2003robust}. Inspired by this work, future research will focus on developing robust extensions of matrix-variate factor analysis \citep{wang2019-mfa} and bilinear factor analysis \citep{zhao2024-bfa,ma2024-rbfa} using the MMCD estimates, with potential applications in macroeconomics, such as constructing robust international trade networks \citep{chen2022-modeling}. Given the high robustness of the MMCD estimates, its extensions to robust matrix-variate linear discriminant analysis and canonical correlation analysis are also under investigation. Furthermore, since HRFPCA is specifically designed for matrix-valued data, an important future direction is its extension to tensor-valued data.

\bibliography{journall,jhzhao-pub,lit,hrfpca}
\bibliographystyle{elsarticle-harv}
	
\end{document}